\documentstyle[preprint,aps]{revtex}

\begin{document}
\author{Helio V. Fagundes and Evelise Gausmann}
\address{{\it Instituto de F\'{\i}sica Te\'{o}rica, Universidade Estadual Paulista, }%
\\
{\it \ Rua Pamplona, 145, S\~{a}o Paulo, SP 01405-900, Brazil}}
\title{On Closed Einstein-de Sitter Universes}
\date{revised 18 August 1997}
\maketitle

\begin{abstract}
We briefly summarize the idea of cosmological models with compact, flat
spatial sections. It has been suggested that, because of the COBE
satellite's maps of the microwave background, such models cannot be small in
the sense of Ellis, and hence are no longer interesting. Here we use Lehoucq
et al.'s method of cosmic crystallography{\it \ }to show that these models
are physically meaningful even if the size of the spatial sections is of the
same order of magnitude as the radius of the observational horizon.

PACS numbers: 98.80.Hw, 04.20.Gz

\ 
\end{abstract}

\begin{center}
{\bf I. INTRODUCTION }
\end{center}

Einstein-de Sitter cosmological model (EdS) belongs to the
Friedmann-Lema\^{\i}tre-Robertson-Walker family, with null cosmological
constant and matter density ratio $\Omega _0=1.$ In the case of a
pressureless energy-momentum tensor as the source in Einstein equation,
which represents today's universe, EdS's metric is 
\begin{equation}
ds^2=c^2dt^2-(t/t_0)^{4/3}\left( dx^2+dy^2+dz^2\right) ,  \label{metric}
\end{equation}
where $t_{0\text{ }}$is the present age of the universe in this model.
Usually the spatial sections of EdS are taken to be the infinite Euclidean
space $E^3$, which has a trivial global topology, i.e., it is a simply
connected space. But we may also consider closed (i.e., compact and without
border) Euclidean manifolds $M^3,$ which are related to $E^3$ through the
isometry $M^3\simeq E^3/\Gamma $, where $\Gamma $ is a discrete group of
isometries of $E^3$; they have a nontrivial global topology, with the
fundamental group isomorphic to $\Gamma .$ See Wolf [1] for a rigorous
mathematical development, or Ellis[2] for a more graphical description
(unfortunately the latter contains a mistake, which will be discussed
below). For cosmology the interesting closed Euclidean manifolds (CEMs) are
the six orientable ones, and here we use the labels $E1$ -- $E6$ for them,
as in Lachi\`{e}ze-Rey and Luminet [3], hereafter LaLu. These CEMs can be
obtained by identification of pairs of faces of a cube ($E1$ -- $E4$) or a
rectangular prism with a regular hexagon as basis ($E5$, $E6$). Such a
polyhedron is called a {\it Dirichlet domain, }or {\it fundamental polyhedron%
} (FP) for the manifold. The FP with identified face pairs represents
comoving space, and is the real space where the sources are. The elements of 
$\Gamma $ act upon the FP to generate its replicas in $E^3$, which is the
locus of the repeated images of the sources in the FP. See, for example,
LaLu's review.

Ellis and Schreiber [4], hereafter E\&S, presented the idea that, with a
universe not only closed but also small, we might have an alternative to the
inflationary scenario as an explanation for the large scale uniformity of
the distribution of matter in the universe. Their point is that, if the
dimensions of the FP are of the order of a few hundred megaparsecs, then one
would not need an actual homogeneous distribution in real space, the
observed large scale uniformity being the result of repeated images of the
sources in the FP.

Recently a number of authors - Sokolov [5], Stevens, Scott, and Silk [6], de
Oliveira-Costa and Smoot [7], de Oliveira-Costa, Smoot, and Starobinsky [8],
and others - have argued that the maps of the cosmic microwave background
radiation produced by the NASA satellite Cosmic Background Explorer (COBE),
put large lower limits on the size of the FP, should one adopt a 3-torus $%
T^3 $, which is the simplest CEM ($E1$ in LaLu and here), as a model for
cosmic space. Let Hubble's constant be $H_0=100h$ km s$^{-1}$Mpc$^{-1}$,
where $0.4\leq h\leq 1.0$, and $R_H=2c/H_0=6000h^{-1}$ Mpc be the radius of
the observational horizon. If we take a cube with edge of comoving length $L$
as the FP, then the smallest of these limits known to the authors is the one
obtained in Ref. [5], $L\geq 0.7R_H,$ and the largest is that of Ref. [7], $%
L\geq 1.2R_H.$ Because the dimensions of the the FP turn out not to be small
in E\&S's sense, Refs. [6--8] imply that models with nontrivial topology are
no longer interesting.

But these results have been obtained by harmonic analysis of COBE maps,
tacitly assuming that each spot on the surface of last scattering (SLS)
should be an image of a density fluctuation with diameter smaller than that
of the FP. As Roukema [9] has argued, a fluctuation might cross the border
of the FP, with the net result that its extension would be larger than the
diameter of a single cell intersecting the SLS. For example, if we think of
a fluctuation that ``spirals'' a number $n$ of times around the 3-torus,
then those lower limits for $L$ would have to be divided by $n$ and we might
recover a small universe in the sense of E\&S.

The above point deserves a more detailed study. Meanwhile, in this paper we
adopt the large limits as valid, to argue that even then one can predict
observable effects for such models, which of course are not small universes.
The point is that the multiple connectedness of their topology has its own
consequences, even if their size is a large fraction of the observable
universe's volume. We use the method proposed by Lehoucq, Lachi\`{e}ze-Rey
and Luminet [10], hereafter LeLaLu, based on a plot of distances between
cosmic images. Because of the periodicities involved in CEMs, one obtains
sharp peaks in this distribution, while the prediction for an infinite
universe (or a finite one containing the {\it whole }observable region)
presents a distribution without such peaks.

\begin{center}
{\bf II. COSMIC CRYSTALLOGRAPHY}
\end{center}

As discussed in Ref. [2], space sections of cosmological spacetimes are
orientable, so the CEMs referred to in this paper are the orientable ones.
There are six families of these, as classified by their global topology.
Their rigorous mathematical study is found in [1]; Ellis [2] interpreted
Wolf's esoteric expressions in colloquial terms, as summarized in E\&S's
Table 2. Unfortunately type $T=4$ in this table, which we call $E4$, is
described as resulting from the identification of opposite sides of a cube
with ``all pairs rotated by $180^{\circ }$.'' But, as pointed out by Bernui
et al. [11], this prescription generates the projective space $P^3$, which
does not admit a Euclidean metric. A correct pictorial description was
obtained by Gomero [12], and here we use a modified version of his result:
With a coordinate system $(x,y,z)$ with origin at the FP's center and axes
perpendicular to its faces, a set of face-pairing generators for group $%
\Gamma $ in $E4=E^3/\Gamma $ is $\{a,b,c\}$, with

\begin{eqnarray}
a(x,y,z) &=&(x+L,\ -y,\ -z)\text{ ,}  \nonumber \\
b(x,y,z) &=&(-x,\ z+L,\ y)\text{ ,}  \label{abc} \\
c(x,y,z) &=&(-x,\ z,\ y+L)\text{ ,}  \nonumber
\end{eqnarray}
which satisfy the relations $a^{-1}ca^{-1}b=caba=b^{-2}c^2=1.$

As for the other five CEMs, they are described by their face pairings in
LaLu, Table 17, as $E1-E3$, $E5$, $E6.$ (Their $E4$ is based on E\&S's
mistaken rule for $T=4$.) The basic prediction of multiply connected
universes is the formation of repeated images of cosmic sources. But the
detection and recognition of multiple images is not an easy task; see LaLu,
\S 11.2. Among the alternative proposals for discerning a nontrivial cosmic
topology is the {\it cosmic crystallography} idea of LeLaLu, which we
illustrate here by an example: Take an image in $E4$ at point $P=(x,y,z)$
and another at $bab^{-1}(P)=(x-L,-y+2L,-z)$; the square of their comoving
distance is 
\begin{equation}
d^2=5L^2-8Ly+4y^2+4z^2\text{ .}  \label{d2}
\end{equation}
LeLaLu's method is to make a table of the comoving distances between all
pairs of images in a given catalog, then plot $n(d)$, the number of
occurrences of each value of $d$, versus $\left( d/L\right) ^2$. As
suggested by Eq. (3), such plots should have peaks on integral numbers,
which will depend on the particular topology and the motions separating the
images. Real catalogs are not yet deep enough to reveal these peaks,
especially if $L$ is very large. But LeLaLu do several simulations, for the
six CEMs, with $L=1500h^{-1}$Mpc, a pseudo-random distribution of $50$
sources in the FP, and a catalog depth of $Z=4$. We might think of these
sources as galaxy clusters or superclusters, in a `{}`top-down{}'{}' picture
of structure formation, or as galaxies and protogalaxies in the now favored
`{}`bottom-up{}'{}' picture - cf. Peebles [13].

We have done other simulations, with our present intention of showing an
effect of a nontrivial space topology on a {\it large }universe. Our sources
include our Galaxy at the observer's position (0,0,0), the other ones being
at pseudo-randomly chosen positions in the FP. (Strictly, except in case E1,
which is homogeneous in the geometric sense, putting the observer at the
FP's center is un-Copernican; but it does not make much difference in the
present study.) And we supposed images distributed in all of the observable
universe (depth $=R_H$, hence highly unrealistic at present; but not
unimaginable, given the progress in such powerful observational tools as
gravitational lensing and the Lyman alpha forest of quasar absorption
lines), presuming that the precursors of today's structures have been
present, at least in embryonic form, from recombination time onwards; see
also Fagundes [14]. Then we calculated and plotted $n(d)$ for both $L=0.7R_H$
(20 sources) and $L=1.2R_H$ (40 to 101 sources). For $L=4200h^{-1}$ Mpc our
plots for $E1$ -- $E3$ agree with LaLeLu's in their common range. For $E5\,$
and $E6$, we got plots different from theirs, because we chose as our FP a
hexagonal prism with different dimensions: in our case the hexagon's
shortest diameter, not its side, is equal to $L$, the length of the vertical
edges.

Fig. 1 shows our both plots for the corrected $E4$ universe and for our
version of the $E5$ case. There are four neat spikes on integral numbers for
the smaller universes, and one for the larger $E5$. Model $E4$ does not show
a peak at $d=L$, because the effect of a translation $L$ in each generator
is strongly masked by the accompanying rotations - cf. Eqs. (2). The larger $%
E4$ does not show a significant peak at $d=L\sqrt{2}$ either; this is almost
twice the horizon's radius, and few ocurrences of this distance are
expected. In this case the plot looks like one for an ordinary (infinite)
EdS model, and we would have to look for other indicators of a nontrivial
topology - see below.

We have not made simulations for the asymmetric $T^3$ models of Ref. [8].
But it should be clear that, for example, in their model $T^1$ with $L_1$ $%
=3000h^{-1}$ Mpc there will be many pairs separated by distances $L_{1\text{ 
}}$and $2L_1$. The reader can convince herself or himself of this through a
simple sketch of five cells in a row.

\begin{center}
{\bf III. FINAL REMARKS}
\end{center}

It is true that large ($L/R_H\sim 1$) closed models do not solve the
homogeneity problem. We may still have inflation, as admitted by E\&S even
for their small models. Actually, most research on cosmic global topology
has been unconcerned with explaining homogeneity - see LaLu for a review;
two recent examples are Ref. [5], where the limit on $L$ was obtained from a
consideration of inflation theory, and Jing and Fang [15], who find $%
L\approx 0.8R_H$ for an $E1$ universe as an explanation for a possible
infrared cutoff in quantum field theory.

The predictions of cosmic crystallography may eventually become testable.
Other recent suggestions for verifying multiple connectedness are Cornish,
Spergel, and Starkman's [16] ``circles in the sky,'' and Roukema's [9]
probabilities for finding repeated images of groups of quasars. On the
theoretical side, the compactness of space is called for by quantum
cosmology; cf. Hartle and Hawking [17], or Zeldovich and Starobinsky [18].
So, even if CEM universes cannot be small (which is by no means certain; see
Introduction), it makes very much sense to continue to explore their
cosmological possibilities and, more generally, those of the rich class of
compact 3-manifolds with a locally homogeneous geometry - see Refs. [19 --
21], for example.

One of us (E.G.) thanks Conselho Nacional de Desenvolvimento Cient\'{\i}fico
e Tecnol\'{o}gico (CNPq - Brazil) for a scholarship. H.V.F. is grateful to
Jeff Weeks and Germ\'{a}n Gomero for conversations and correspondence, and
to CNPq for partial financial support.

---

[1] J. A. Wolf, {\it Spaces of Constant Curvature} (McGraw-Hill, New York,
1967)

[2] G. F. R. Ellis, Gen. Relat. Gravit.{\bf \ 2}, 71 (1971)

[3] M. Lachi\`{e}ze-Rey and J.-P. Luminet, Phys. Rep. {\bf 254}, 135 (1995);
LaLu

[4] G. F .R. Ellis and G. Schreiber, Phys. Lett. A{\bf \ 115}, 97 (1986);
E\&S

[5] I. Y. Sokolov, JETP Lett.{\bf \ 57}, 617 (1993)

[6] D. Stevens, D. Scott, and J. Silk, Phys. Rev. Lett. {\bf 71}, 20 (1993)

[7] A. de Oliveira-Costa and G. F. Smoot, Astrophys. J. {\bf 448}, 477 (1995)

[8] A. de Oliveira-Costa, G. F. Smoot, and A. A. Starobinsky, Astrophys. J. 
{\bf 468}, 457 (1996)

[9] B. F. Roukema, Mon. Not. Roy. Ast. Soc. {\bf 283}, 1147 (1996)

[10] R. Lehoucq, M. Lachi\`{e}ze-Rey, and J.-P. Luminet, Astron. and
Astrophys. {\bf 313}, 339 (1996); LeLaLu

[11] A. Bernui, G. Gomero, M. J. Rebou\c {c}as, and A. F. F. Teixeira,
report CBPF-NF-027/97 of Centro Brasileiro de Pesquisas F\'{\i}sicas

[12] G. I. Gomero, personal communication (1996)

[13] P. J. E. Peebles, {\it Principles of Physical Cosmology} (Princeton
University, Princeton, 1993), Chapter 15

[14] H. V. Fagundes, Astrophys. J. {\bf 470}, 43 (1996)

[15] Y.-P. Jing and L.-Z. Fang, Phys. Rev. Lett. {\bf 73}, 1882 (1994)

[16] N. J. Cornish, D. N. Spergel, and G. D. Starkman, preprint gr-qc/9602039

[17] J. B. Hartle and S.W. Hawking, Phys. Rev. D{\bf \ 28}, 2960 (1983)

[18] Y. B. Zeldovich and A. A. Starobinsky, Sov. Astron. Lett. {\bf 10}, 135
(1984)

[19] H. V. Fagundes, Phys. Rev. Lett. {\bf 54}, 1200 (1985); Gen. Relat.
Gravit. {\bf 24}, 199 (1992)

[20] T. Koike, M. Tanimoto, and A. Hosoya, J. Math. Phys. {\bf 35, }4855
(1994); J. Math. Phys. {\bf 38}, 350 (1997)

[21] J. R. Weeks, {\it The Shape of Space} (Marcel Dekker, New York, 1985)

\newpage\ \ 

Figure legend:\vspace{1.0in}

FIG. 1. Distribution of comoving distances between images in simulations
with 20 (smaller $L$) or 101 sources. The number of images is about the same
(246 -- 280) in each case. The bins have width 0.01.

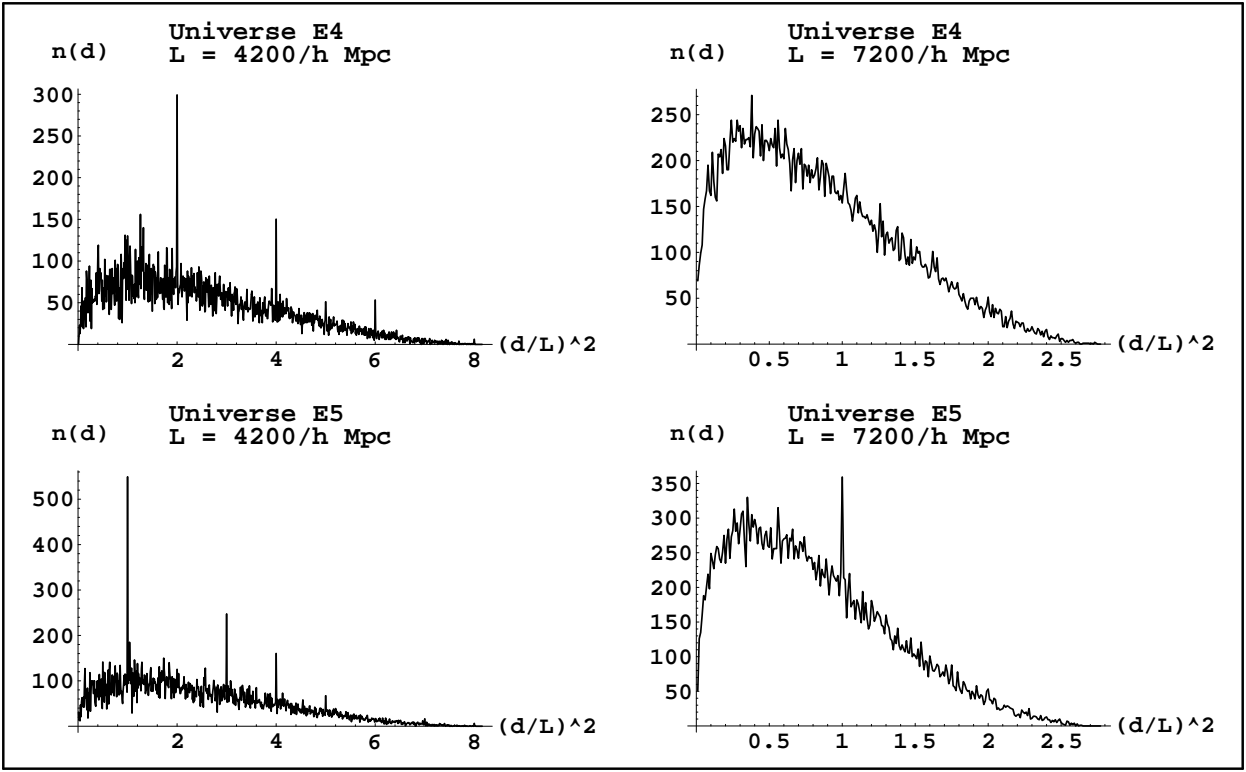
\end{document}